# Pointing task on smart glasses: Comparison of four interaction techniques

Ilyasse Belkacem[a,], Isabelle Pecci[a], Benoît Martin[a]

[a]*LCOMS, Université de Lorraine, France*

**Abstract**

Mobile devices such as smartphones, smartwatches or smart glasses have revolutionized how we interact. We are interested in smart glasses because they have the advantage of providing a simultaneous view of both physical and digital worlds. Despite this potential, pointing task on smart glasses is not really widespread. In this paper, we compared four interaction techniques for selecting targets : (a) the Absolute Head Movement and (b) the Relative Head Movement, where head movement controls the cursor on smart glasses in absolute or relative way, (c) the Absolute Free Hand interaction, where the forefinger control the cursor and (d) the Tactile Surface interaction, where the user controls the cursor via a small touchpad connected to smart glasses. We conducted an experiment with 18 participants. The Tactile Surface and Absolute Head Movement were the most efficient. The Relative Head Movement and Absolute Free Hand interactions were promising and require more exploration for other tasks.

## 1. Introduction

Smart glasses are wearable devices that embed electronic components such as a display screen, an integrated processor, and sensors. They allow visualization and interaction with an interface that is displayed through the lens. Smart glasses provide users with a simultaneous view of both physical and digital worlds. With a constant delivery of information close to the eyes, smart glasses provide a framework for many applications, including medical documentation, surgery, educational functions and gaming [1].

Smart glasses are not only a new screen: they are a new way to interact. The challenge is to provide interaction techniques suitable to use with smart glasses and that can facilitate complex tasks.

As existing mobile interfaces are primarily designed for smartphones, it is difficult to find input interactions dedicated to smart glasses. At the launch of the first commercial

*Email addresses:* `ilyasse.belkacem@univ-lorraine.fr` (Ilyasse Belkacem), `isabelle.pecci@univ-lorraine.fr` (Isabelle Pecci), `benoit.martin@univ-lorraine.fr` (Benoît Martin)

smart glasses, Google Glass [2], the main input interaction technique was voice recognition. However, this technique is inappropriate in shared environments, can be noisy [3], and is less preferred compared to body gestures and handheld devices [4]. Today's smart glasses offer a multitude of sensors and controllers that can be used for input interaction. For example, the GPS and accelerometer can recognize the movement and position of the user to improve their interface experience. As the gyroscope measures the angular velocity and the orientation of the head, the user can move his or her head to interact with the smart glasses application. In addition, a camera embedded on smart glasses could recognize gestures (e.g. hand gesture) to provide input interactions.

In this paper, we compared four smart glasses interaction techniques to control the cursor according to their performance and usability. Two techniques are based on head movement, one is based on hand movement, and one based on a tactile surface. To this end, we first present why pointing with smart glasses is so important by describing our user context. Then, we present research related to interaction techniques with smart glasses. Next, we describe the four techniques developed and their implementation in our prototype. Finally, we present the experiment used to evaluate the four interaction techniques and conclude with an analysis of the results.

## 2. Contribution statement

Smart glasses are considered a rudimentary product due to major constraints such as input methods [5]. The main contribution of this paper is to show evidence that some input methods are easy to use and that smart glasses are not as rudimentary as previously believed. To achieve this, we evaluate user performance of different smart glasses pointing techniques and provide in-depth results through ISO standard following the multidirectional tapping test.

The choice of techniques, head movement and free hand movement, was motivated by the healthcare context and the benefit of not requiring users to hold a device. We investigate the potential of these techniques compared to the tactile surface interaction technique commonly used to point with smart glasses.

## 3. Healthcare context

In healthcare context, the use of health information technology to improve patient care, reduce costs, and increase effectiveness of health professionals is of great interest [6]. Patient records applications on smartphones and tablets reduce costs and improve efficiency [7] in a clinical setting. However, they are not frequently used [6] because they do not allow clinicians hands to be free for medical tasks. Smart glasses can overcome this limitation because they can be used in hands-free mode. Several studies have examined the use of smart glasses in managing health records and reducing the time spent interacting, and many technological studies have shown that smart glasses have potential areas of application in the medical industry [6, 8, 9]. These areas include surgery, telemonitoring (vital sign monitoring and input), electronic health records, telemedicine and medical education, including hands-free photo and video documentation, rapid diagnostic, test analysis and live broadcasting [8]. Healthcare professionals reported that smart glasses improved focus on tasks by reducing movements (which would be required to view multiple remote monitors) and improve situational awareness [8]. In surgery, the display of smart glasses in front of the eyes allow surgeons keep an eye on the operative field [6].

If we look at the interfaces of the existing health records, such as LifeLine [10], Miva [11], and other commercially available systems, a user moves a cursor to select information from a 2D interface in order to have more details [12]. The challenge for smart glasses is to provide suitable pointing interaction techniques in this context.

*3.1. Motivating scenario*

To understand the problem of using the electronic health record, we describe a design scenario that illustrates the typical daily tasks of a doctor with smart glasses.

At the hospital where Dr. Bob works, an Electronic Health Record (EHR) system is being introduced to replace paper health records. The EHR allows different users (health professionals, patients, etc.) to access useful information that contributes to patient monitoring, such as history, allergies, etc.

In addition to the EHR system, Dr. Bob can use smart glasses to display health record information (vital signs, medical imagery, laboratory results, antecedents, etc.) in his field of vision to allow him to have the right information at the right time while staying engaged with the patient. Dr. Bob finds the technology very exciting because it allows him to quickly see the change of condition of patients and react quickly.

Dr. Bob needs to navigate the patient health record displayed on his smart glasses screen by using a pointing technique for various operations: select information for more details, select a character in the virtual keyboard, or select a button or menu on the patient record interface. The pointing can be done by a cursor that Dr Bob controls with a touchpad connected to the smart glasses.

Dr. Bob needs to interact effectively and efficiently. He can view the patient record while standing in the examination room, walking in the hallways of the hospital, or sitting in front of his patient. However, he finds that the pointing technique with the touchpad is easy but difficult to use when doing other activities. A hands-free interaction would be very useful for him because his hands are often occupied during the medical practice (holding medical equipment, palpation, surgery). He previously used another smart glasses that embedded a tactile surface on the branch to allow hand free interaction, but he had to sterilize his hands every time he touched the pointing device, which was inefficient. Hands-free interaction methods are promising for Dr. Bob because they allow him not to interrupt his medical procedure or to waste time sterilizing his hands.

The proposed scenario shows that smart glasses could be integrated in a hospital context provided that the pointing techniques are not binding. Our work is to identify and evaluate hand-free techniques.

# 4. Related works

Several previous research studies on the input interactions with smart glasses are present in the literature. They can be classified into three groups [13]:

- Handheld devices. The user interacts with a handheld controller such as a smartphone or a trackpad.

- Touch. The user taps on parts of his body or wearable devices such as smart rings or smartwatches.

- Non-touch. The user does not need a surface to interact. This group includes interaction techniques like gestures in the air, movements of the head or body, or voice recognition.

The first group includes input interactions with a tactile surface. We chose this type of interaction as the baseline for our study because it is common on mobile devices and is also available with some smart glasses. The third group, non-touch interaction techniques, is unique because it is the least invasive for user. As users already wears smart glasses, we did not want to add another device that could make mobile interactions difficult to complete. We focus the related work on this non-touch group of interactions, in particular interactions using hand and head movement.

## 4.1. Head movement interaction

Head-tracking interfaces provide a way to interact with devices through the movements of the head [14]. Two different kinds of technologies are reported within the references: camera-based and sensor-based. The camera-based approach [14, 15, 16, 17, 18, 19] uses a face-tracking system that interprets human head movements in real time. It has more advantage on interfaces that are not mounted on the head (laptop, tablet) to avoid additional sensors or equipment worn on the head.

A recent study [14] explains the use of camera approaches and the fact that personal computers have become more powerful. In addition, camera systems for computers are now readily available, which has contributed to advances in computer vision. In their work, the authors use the front camera on a tablet to capture the movement of the head. This approach suffers from the technical limitations of capturing motion in a scene with changing the lighting.

The second 'sensor-based' approach is less intrusive and allows for accurate measurement [20]. This approach is used in several applications for controlling an stationary interface of a computer for the disabled [21, 22, 23].

Head tracking with sensors is more natural with Head Mounted Displays (HMDs), because they already have sensors to track the movement of the head. For virtual reality environments, the movement of the head allows designers to identify the perspective view of the user and plan the virtual environment as it is experienced in the game [24].

The movement of the head on HMDs is used to perform head gestures to authenticate [3] or to control a Pac-Man game on a large projection on the opposite wall. Users control all four head movements (up, down, left, right) to direct the Pac-Man, but no precision is needed [25]. Researchers have mostly focused on the development of new tracking methods and their integration into prototypes, but few analyzed human performance.

Yu et al. [26] investigate head-based text entry for HMDs by controlling a cursor on a virtual keyboard using head rotation. The study only evaluates the user's performance for text entry, not the pointing task. The study did not compare the relative head movement to the absolute one. There have been other previous user performance analyses of head orientation on smart glasses [27, 28, 29] and in a virtual reality environment [30], but results are quite different. Jalaliniya et al. [27] compares three modalities (mouse, gaze and absolute head movement). Their results show a better speed for gaze pointing but with less precision compared to the movement of the head. Conversely, in a virtual reality environment, Qian et al. [30] found that absolute head movement is better than gaze movement. The two studies are difficult to compare because experiences are not described at the same level of detail and contexts differ (real environment vs virtual one, smart glasses vs no smart glasses). [28, 29] combined the movement of the head with gaze movement. Even if results are interesting with this combination, a user in a pointing task interaction would have to keep his eyes on the target for a long time to validate, due to the eye gaze technique. In a healthcare context, the doctor would lose contact with his patient for too long of a time. For these reasons, we ruled out this interaction technique in our study.

Unlike these evaluations, here we evaluated two techniques of head movement, relative and absolute, by providing in-depth results through the ISO standard and Mackenzie accuracy measures (see Results section).

Recently, the glasses developed by Microsoft HoloLens [31] adopted the technique of ray casting as a standard way to navigate in the virtual world and manipulate virtual objects. Although the name of the technique used - Gaze - refers to the direction of the user's eyes, this technique uses the absolute position and orientation of the user's head to determine his gaze vector. This vector gives a direction for the ray. A cursor appears on the intersection of this ray with the object closest to the user. The ray casting technique is typically used in a 3D environment with a 360° rotation of the head, which is not appropriate for our scenario. As the doctor is facing a patient, it is not appropriate to completely rotate the head to interact with the glasses. The doctor must be able to move his or her head without losing his interaction with the patient. Thus, the technique of ray casting as it is proposed in 3D is too complex for our context but could be adapted to use in some absolute movements of the head.

*4.2. Free hand interaction*

Free hand interactions need sensors to capture the movement of the hand, which is usually a camera. [32, 33, 34] use the RGB camera and discuss the feasibility of their prototypes through vision-based methods. We provide a technical evaluation to evaluate the accuracy of the hand tracking, but the results are not efficient enough.

Other types of cameras are prototyped to have a higher precision, such as CyclopsRing [35] which proposes a ring style portable device with fisheye imaging, or Digits [36] which uses two small infrared cameras placed on the wrist to detect the full 3D pose of the hand. A precision study was carried out which shows a significant error rate. On the other hand, ShoeSense [37], which uses a Kinect depth camera placed on the shoes, demonstrates the best accuracy of all the systems mentioned above.

In terms of the use of free hand, Huang et al. [38] proposes Ubii to facilitate the interaction between the physical world (office objects) and the digital world. The user's gestures are captured by the smart glasses to allow the individual to control the objects in his environment. WeARHand [39] proposes manipulation of 3D objects in augmented reality with the hand. The manipulation of objects with the hand is often used with virtual reality helmets because it allows users to enforce the immersion with realistic gestures according to Sait, et al., [40] which demonstrates the usability of the manipulation in a virtual game environment with the hand. Hand gestures can be used to perform commands [41, 42].

Commercially available HMD like Hololens [43] and Meta 2 [44] offer this type of interaction to interact in augmented reality applications. While the works cited deal with the interaction with the free hand, none addresses the problem of controlling the cursor on smart glasses with the hand movement. We attempt to examine this aspect by studying performance and usability.

## 5. Techniques

We present four interaction techniques for pointing tasks on smart glasses. As describe above, we wanted to design interaction techniques that could be used by a doctor while he visits a patient. Based on related work, we selected the head movement and free-hand techniques. We first describe our baseline technique that use tactile surface interaction. Then we present three interaction techniques adapted to our context: Absolute Head Movement, Relative Head Movement and Absolute Free Hand.

### 5.1. Tactile Surface (TS)

The concept of interacting on tactile surface is common to many smart glasses [2], but the location of the surface varies. The tactile surface can be moved onto a specific device connected to smart glasses, as in Figure 1, or embedded on the branch of the smart glasses. The technique of the TS on smart glasses works like the touchpad on a laptop. Whatever the implementation on the smart glasses, localized or not, the users hand is solicited. We chose to use smart glasses that provide a remote touch surface on a controller (Figure 1). The user takes the controller in his hands and directs the cursor by moving a finger on the tactile surface. The faster the fingers speed, the greater the movement of the cursor is. The position of the cursor is relative.

### 5.2. Head Movement (AHM and RHM)

The head movement is an intriguing technique because it allows users to perform tasks in a mobile context even if the hand is occupied. We have developed two techniques: AHM and RHM. These two techniques allow the user to control the cursor without contact by moving his or her head. At the beginning, the cursor is placed at the center of the screen. This position matches the position of the users head. To move the cursor, the user must rotate his head. The movement of the cursor is in the same direction of the movement of the head. The rotations used to move the cursor on the screen are "Pitch" and "Yaw" (see Figure 2). The "Yaw" rotation matches the horizontal movement of the cursor and the "Pitch" rotation matches the vertical movement of the cursor. The position of the cursor on the screen differs by the technique used, AHM or RHM.

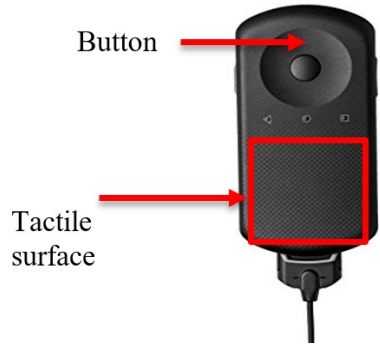

Figure 1: Tactile surface on smart glasses

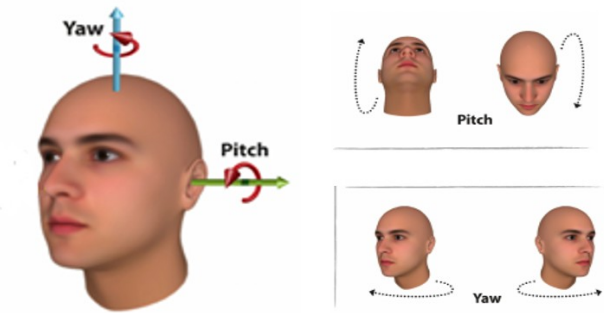

Figure 2: Illustration of pitch and yaw movement [45]

*5.2.1. Absolute Head Movement (AHM)*

This technique is based on the angular orientation of the head. The angular orientation (0°, 0°) is associated with the initial position of the head, which matches the position of the cursor on the center of the smart glasses screen. If the user wants to change the initial orientation of his head, he can press a button to calibrate the cursor.

The position of the cursor is absolute, that is, if the user moves his head 10° to the east at two speeds, v1 or v2 where v1 = v2, the cursor moves to the same position. A constant angular orientation always matches to the same cursor coordinates.

The maximum rotation required to reach the end of the screen is a horizontal movement of 20° East (or West) or a vertical movement of 10° North (or South) as shown in Figure 3.

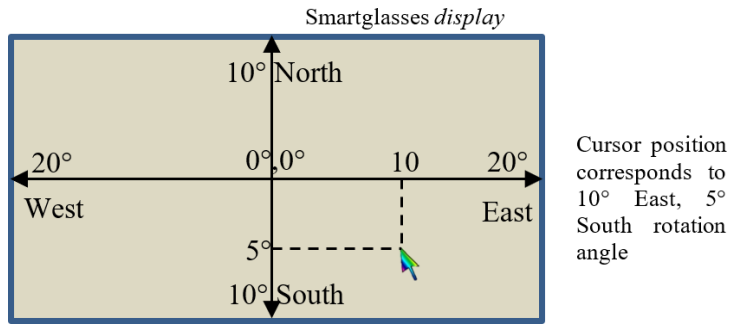

Figure 3: Maximum rotation degrees for AHM technique

*5.2.2. Relative Head Movement (RHM)*

The cursor moves on the screen according to the speed of rotation of the head (angular velocity in rad/s). The higher the speed is, the greater the movement of the cursor is. When the head moves, the cursor starts from the last position on the screen and uses the speed of head movement, i.e. if the user moves his head 10 East with two different speeds, v1 and v2 (v1 = v2), the cursor does not move to the same position. If the user cannot control his speed during RHM, the system cannot remain calibrated. A threshold of 0.04 rad/s is applied to ignore slight movements below this level that may potentially be noise and not indicative of real motion.

If the user wants to change the initial orientation of his head, he can calibrate the cursor and return it to the center using two movements in two opposite directions with two different angular velocities $\omega_1$ and $\omega_2$ as illustrated in Figure 4. This mechanism is an advantage, because it allows the user to recalibrate the cursor in a mobile context in which the users head orientation might frequently change - even if hands are occupied. On the other hand, it requires mental effort and time to complete the movement. A button is also available for this function.

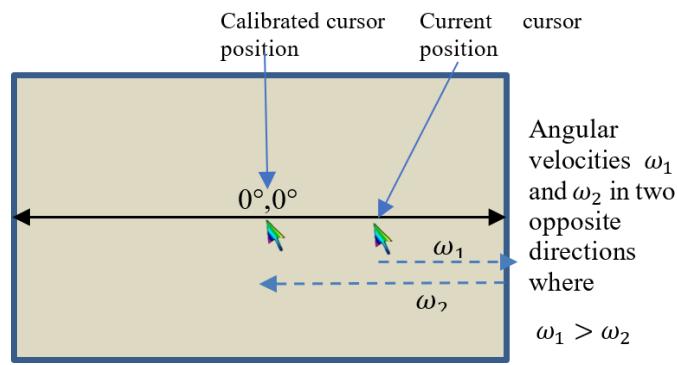

Figure 4: Cursor calibration mechanism for the RHM

*5.3. Absolute Free Hand (AFH)*

The absolute free hand (AFH) technique allows the user to move the cursor with his index finger. To avoid hand-tracking problems during the evaluation, we chose a depth camera with precise tracking of hand movement. A depth camera was installed on smart glasses as shown in Figure 5. The user must place the hand in an area close to the smart glasses - approximately two feet (45cm) for the best performance.

The position of the index finger is captured by the camera and controls the translation of the cursor position on the screen. The cursor moves on the screen following the movement of the index finger. The movement is absolute, i.e. the fixed position of the index finger in the real world always matches the cursors coordinates on the screen. A Control-Display ratio of 1 means that the movement of the finger in real world is represented exactly in the smart glasses display device (i.e. one-to-one movement). This choice is supported by measured accuracy and user feedback that a one-to-one movement felt the most natural [46]. Precision and naturalness are crucial for our scenario.

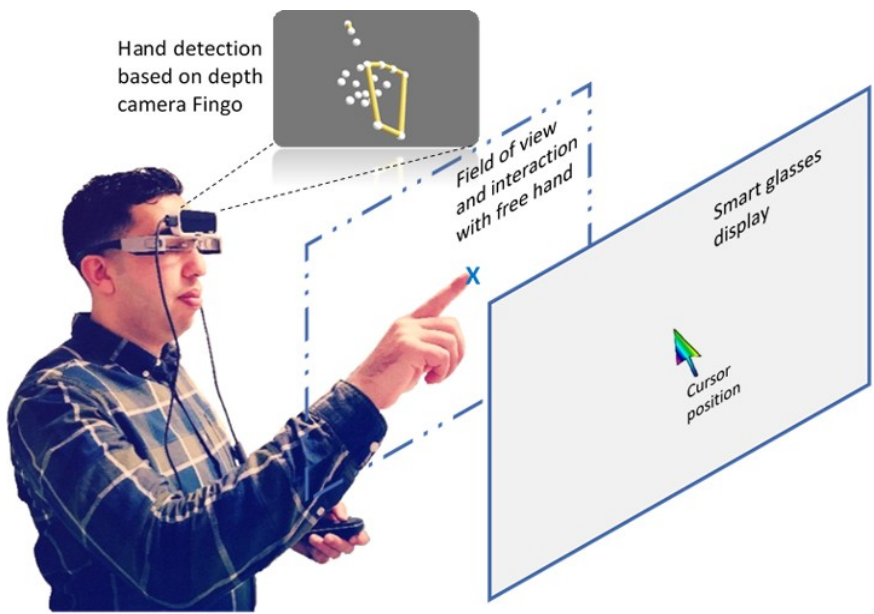

Figure 5: Hand tracking and cursor positioning on smart glasses display

## 6. Experiment

This experiment aims to evaluate three pointing techniques on smart glasses (Absolute Head Movement (AHM), Relative Head Movement (RHM) and Absolute Free Hand (AFH)) in terms of performance and comfort, compared to a baseline (Tactile Surface (TS)).

## 6.1. Evaluation of pointing devices and techniques performance
### 6.1.1. Fitts' law and ISO 9241-411 evaluation standard

Before presenting our methodology, we present Fitts' law and the ISO 9241-411 used to compare the four pointing techniques. Fitts' law [47] is one of the fundamental laws to evaluate a pointing task. It indicates that the time required for a user to move a cursor to a target area is based on the ratio of the distance to the target and the width of the target. The longer the distance is, the smaller the target size is, and the longer it takes. Fitts defines the index of difficulty (ID) depending these two metrics (distance $D$, width $W$). We chose to use the effective index of difficulty (IDe) [48, 49] depending on the effective distance and width the user has performed.

Linear regression defines the linear relationship between the movement time $MT$, and the index of difficulty $ID$. The equation of the line is given by:

$$MT = a + b \times ID_e \qquad (1)$$

where $MT$ is the completion time between the starting point and the point of arrival on the target. The least squares method is used to find the parameters of intercept ($a$) and slope ($b$). This equation is called the equation of Fitts' law.

Speed and accuracy are the main measures to compare two techniques or devices. The speed is usually represented by the ($MT$) and accuracy by the error rate (the percentage of target selections when the cursor is out of the target). These measures are generally analyzed over a wide representative range of indices of difficulty, $IDe$ (with different widths $We$ and distances $De$) [50].

ISO 9241-411 "Evaluation methods for the design of physical input devices" [51] specifies the quality of the device or non-keyboard input technique in terms of performance criteria through various performance tests: One-directional tapping test, Multi-directional tapping test, Dragging test, etc. Among these performance tests, we use the multi-directional tapping test to evaluate pointing movements in many different directions. This task involves positioning the cursor within a target area displayed on the screen. Discs are spread over a circle. A disc becomes a target when its color changes. The goal is to select each target presented to convert all disks to become targets. Two consecutive targets are diametrically opposed. The participant can control the cursor using a pointing technique. The first target is not taken into account because it is only used to start the test. A 9-target multi-directional selection task with target order is shown in Figure 6.

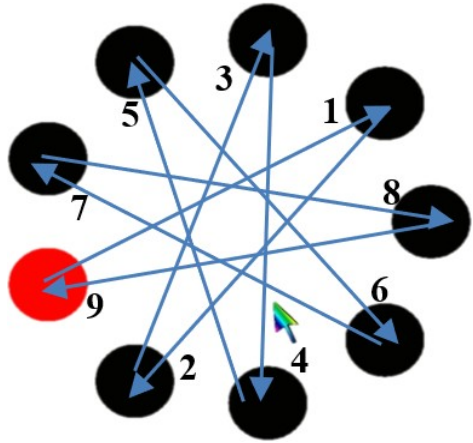

Figure 6: ISO 9241-411 multi-directional position select task with the order of the targets, the red disk represents the first target

Throughput ($TP$), also known as a performance index, is the fundamental and global metric quantifying the performance of a device or input technique in this ISO standard. This measure combines the effects of parameters ($a$) and ($b$) of the regression model into an easily comparable measure despite criticism that it can replace the regression model [52].

The ISO 9241-411 standard also provides a questionnaire of seven questions to evaluate the comfort and the effort of using the device or the technique in a subjective way on a scale from 1 to 5.

*6.1.2. MacKenzie's accuracy measures*

Unlike previous measures that were based on a single scoring task measure, MacKenzie et al. [53] propose seven measures that study the behavior of cursor movement during a pointing task. In a "perfect" pointing task, the participant points directly to the center of the target (see Figure 7.a). The movement that the user performs is different from that in practice. Figure 7.b shows a target selection task from left to right and the cursor path in practice with five sampling points. The seven measures are:

1. *Target Re-Entry (TRE)* which measures if the cursor enters target region, exits without selecting the target and re-enters another time (see Figure 7.c).
2. *Task Axis Crossing (TAC)* which measures if the cursor crosses the axis of the task (a straight line between the cursor position and the center of the target) (see Figure 7.d).
3. *Movement Direction Change (MDC)* which measures whether the cursor path changes direction relative to the task axis (see Figure 7.e)
4. *Orthogonal Direction Change (ODC)* which measures whether the cursor path changes direction relative to an axis orthogonal to the task axis (see Figure 7.f).
5. *Movement Variability (MV)* which measures whether the cursor path remains on a straight line parallel to the task axis. It is measured by the standard deviation of the points ($x_i, y_i$) of the cursor path relative to the task axis (transformed to $y = 0$)

6. *Movement error (ME)* which is the average deviation of the cursor path points from the task axis where points are above or below the task axis.
7. *Movement offset (MO)* which is the average deviation of the cursor path points from the task axis. Figure 7.g shows a comparison between $MV$, $ME$ and $MO$.

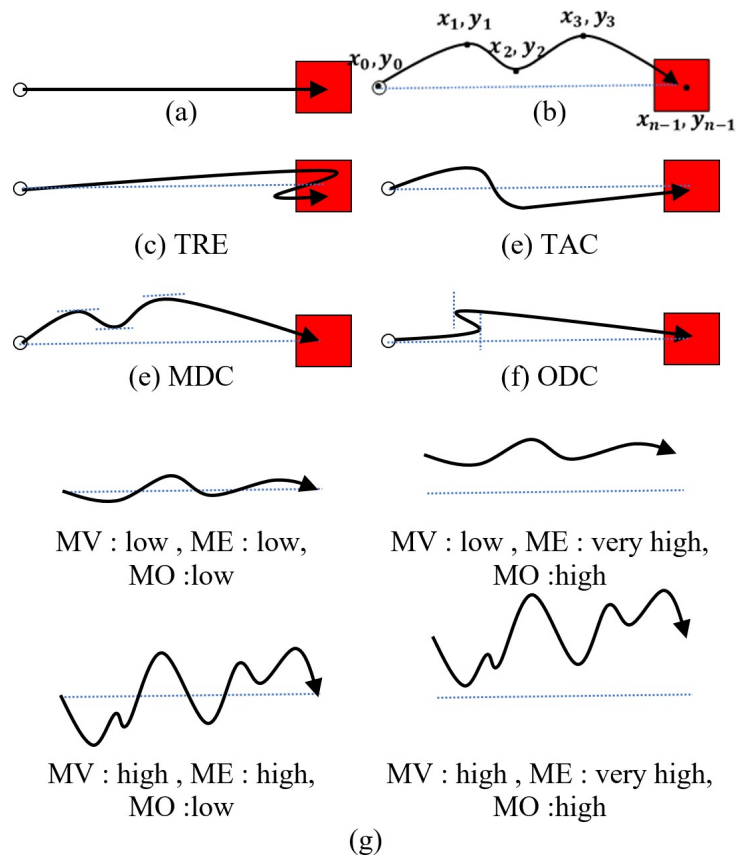

Figure 7: Path variation. (*a*) perfect target selection task (*b*) target selection task in practice (*c*) target re-entry (*d*) task axis crossing (*e*) movement direction change (*f*) orthogonal direction change (*g*) 4 comparisons between movement variability, movement error, and movement offset

*6.2. Task*

Participants performed a multidirectional pointing task of ISO 9241-411 with 9 targets (Figure 6) in 4 configurations mapping to an index of difficulty ID (Figure 8). The cursor was initially on the center of the circle. A disc became a target when it became red. The participant had to move the cursor on the screen which was represented by an arrow.

The pointing task ended with the validation of the final target selection. As the validation task was not the purpose of our study, participants validated the selected target with a button on a handheld device (see Figure 1). After validation, the disk became green if the cursor was really on the target and orange in case of error.

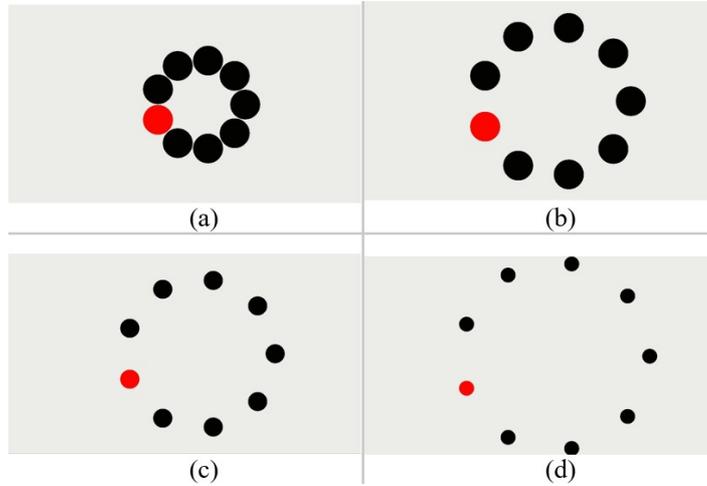

Figure 8: Screenshot of the experimental pointing task with the different sequence conditions (*a*) Sequence 1: *ID* = 2*bits*, *W* = 100*px*, *D* = 300*px* (*b*) Sequence 2: *ID* = 2.58*bits*, *W* = 100*px*, *D* = 500*px*. (*c*) Sequence 3: *ID* = 3.11*bits*, *W* = 65*px*, *D* = 500*px*. (*d*) Sequence 4: *ID* = 3.76*bits*, *W* = 50*px*, *D* = 630*px*.

*6.3. Apparatus*

The experiment was performed using Epson binocular transparent smart glasses. This device offers high image quality with its OLED display (1280× 720 px). The virtual screen is equivalent to a screen of 80 inches to 5 meters.

The smart glasses embed various sensors (GPS, compass, gyroscope, accelerometer, light, etc.). They are connected to a controller through a cable. The controller contains a tactile part used as a trackpad, and various buttons. For the Tactile Surface technique, we used the coordinates provided by the touchpad of the smart glasses. For both head movement techniques (AHM and RHM), it was necessary to capture the degree of head rotation and angular velocity. We used the class associated with glasses sensors: Sensor Manager, Rotation Vector Sensor and Gyroscope Sensor. The smart glasses operate under an Android 5.1 operating system and the test application was developed in Java using the Android SDK.

A Fingo uSens[1] camera was mounted on the edge of the smart glasses (Figure 9) to detect the users hand in the field of vision. The camera was connected to a Macbook Pro (Intel Core i7 2.5 GHz, with 4GB of RAM), which runs under Windows 10 (Fingo does not work on Mac OS). The participant did not have access to the Macbook display. The Macbook was used as a server to transmit the position of the index finger to the smart glasses via the WIFI network. The hand and finger recognition application developed under Unity was launched on the Macbook during the experiment.

---

[1] https://www.usens.com/fingo

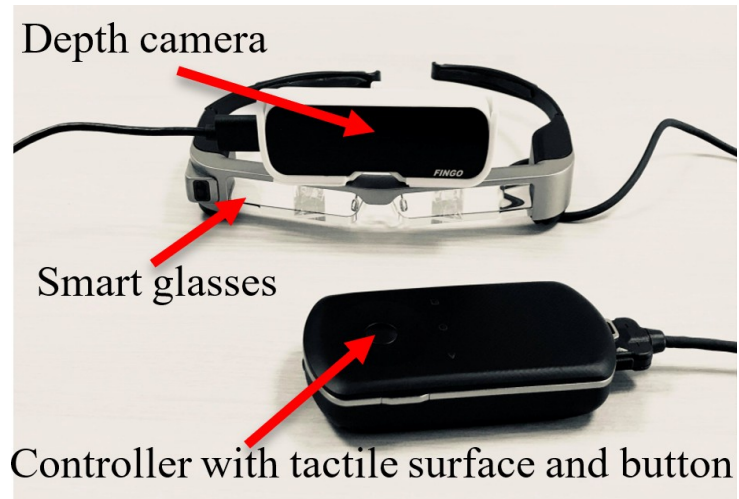

Figure 9: The materiel used in the experience.

Each participant was comfortably seated, wore the smart glasses, and held the controller in his hand. The experiment took place in a room with consistent lighting conditions (closed window shutters and interior lights). Figure 10 shows the configuration of the room.

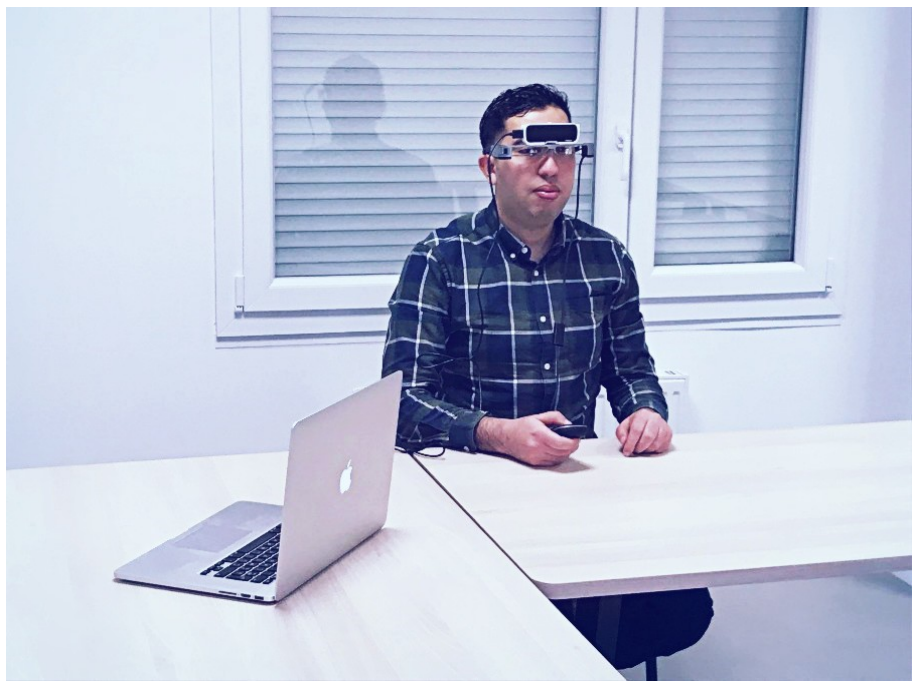

Figure 10: Configuration of test room.

For the absolute free hand technique (AFH), we used the depth camera. This camera was not compatible with our smart glasses processor and provided only an SDK for Unity.

It was necessary to develop a server application on Unity that retrieved the hand position (fingers and bones) in space and sent them to the Android client application on the smart glasses to control the position of the cursor. Figure 11 shows an example of hand recognition on our Unity application.

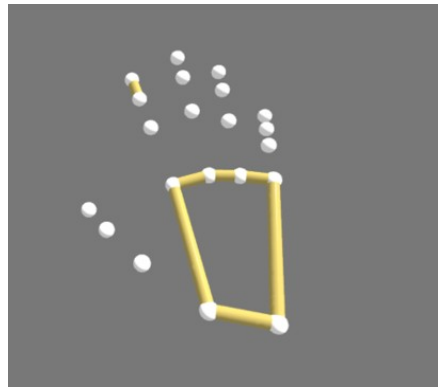

Figure 11: Screenshot from Unity server of hand detection

*6.4. Participants*

We recruited 18 unpaid volunteer participants (6 females), with normal vision. All of them were University students studying Computer Science. Their ages ranged from 20 to 30 years (Table 1). Participants were daily computer users and familiar with the manipulation of the mouse cursor. Seven participants used a touchpad every day, five used it often, and six rarely. Only two participants had previously used the head movement technique to control a game. Five participants had previous experience with a depth camera but were not experts. The smart glasses were a new device for all the participants.

Table 1: Age distribution of participants

| Age range | Frequency |
|---|---|
| [20-25] | 15 |
| [26-30] | 3 |

*6.5. Procedure*

After entering the lab, each participant signed a consent form and was invited to read a document that specified the purpose of the experiment, the task to perform, and the devices to use. They also filled out a pre-questionnaire about age, sex, dominant hand, and useful computer skills for the test (computer and cursor use, touchpad use, use of head movement, the use of the depth camera and the use of smart glasses). Participants began the test with a training session on the four pointing techniques by selecting and validating 8 targets for each technique (Figure 12). The experimenter could help them. The results were not recorded but the problems were gathered.

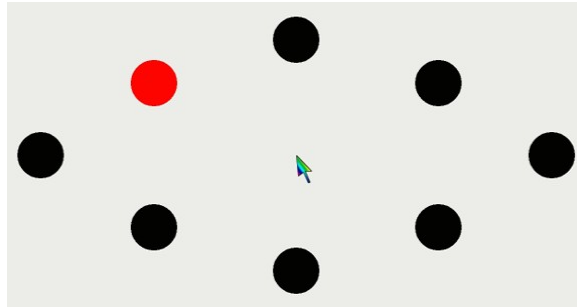

Figure 12: Screenshot of the training session.

After the training session, the application test started. Each participant completed two sessions separated by at least 24 hours and maximum 72 hours, without a training session between. The participant could not receive help during the test. For each session, participants were invited to test each interaction technique (TS, RHM, AHM, AFH). Participants were randomly divided into 4 groups. The order of interaction techniques in each group followed a balanced Latin square to minimize the asymmetrical learning effects from one pointing technique to the other (Table 2). From one session to the next, participants were assigned to a different group. For each interaction technique, the participant performed two blocks. A block was a group of four sequences. For each sequence, participants were asked to select nine target disks of a given configuration as quickly and accurately as possible. Each sequence corresponded to a configuration (an index of difficulty ID). According to the standard, 4 sequences in increasing order of difficulty were presented (see Figure 8).

Table 2: The order of different techniques for each group

| Group 1 | TS  | AHM | RHM | AFH |
|---------|-----|-----|-----|-----|
| Group 2 | AHM | AFH | TS  | RHM |
| Group 3 | AFH | RHM | AHM | TS  |
| Group 4 | RHM | TS  | AFH | AHM |

The session started with a splash screen that asked the participant to press the button on the screen to begin the first technique. When the participant completed the first block with the four sequences, the second block followed. Between two pointing techniques, a screen that separated each technique was displayed. The participant had to press the button used for validation to continue to the next technique. The participant could pause or adjust the cursor at the end of a technique or even at the end of a sequence, since the first target of each sequence was only used to start the test. At the end of the test, participants were asked to complete a questionnaire.

*6.6. Design*

The experiment followed a 4×2 within-subjects design with: pointing technique (tactile surface, absolute head movement, relative head movement and absolute free hand) and the session (1,2) as factors.

The dependent variables were throughput, movement time, and error rate. The index of difficulty ID represented by the sequence (Figure 8) is an additional independent variable that ensures the participant is confronted with a wide representative range of difficulties. However, it is not a primary one since dependent variables were computed by considering IDe (see Fitts' law and ISO 9241-411 evaluation standard section ) which is dependent from user.

*6.7. Collected data*

During this experiment, we recorded each pointing task (selection and validation). The total number of trials was 18 participants × 4 techniques × 2 sessions × 2 blocks × 4 sequences × 9 targets = 10368.

For each task, we recorded the start point, the validation point and the movement time between the two points with the different properties (D, W, block number, session number, and technique.) We also recorded the cursor path points. The data collected were saved in CSV.

We also collected all user's feedback:

- sorting of techniques in preference order;

- usability of each technique via SUS questionnaire;

- evaluation via ISO 9241-411 questionnaire that evaluates performance, comfort, and effort in using computer pointing devices;

- informal comments and observations.

## 7. Results

The experiment lasted 48 minutes on average (26 minutes on average for the first session and 22 minutes on average for the second session).

*7.1. Throughput, error rate and regression model*

Figure 13 shows the throughput (bits per second) by technique and by session. The throughput is 1.82*bps* (session 1 *sd* = 0.37) and 2.02*bps* (session 2 *sd* = 0.28) for the TS technique, 1.72*bps* (session 1 *sd* = 0.32) and 1.99*bps* (session 2 *sd* = 0.30) for the AHM technique, 1.30*bps* (session 1 *sd* = 0.22) and 1.46*bps* (session 2 *sd* = 0.26) for the RHM technique and 1.22*bps* (session 1 *sd* = 0.48) and 1.46*bps* (session 2 *sd* = 0.44) for the AFH technique.

A two-way 4 × 2 RM-ANOVA was carried out on throughput by session and techniques. The throughput in the second session was significantly higher than the first session ($F_{1,136}$ = 14.495, $p < .001$). Also, the main effect of technique was significant $F_{3,136}$ = 28.656, $p < .001$).

Tukey's HSD post hoc tests on techniques were carried out for throughput. Comparisons using the Tukey HSD test indicated that TS and RHM, TS and AFH, AHM and RHM, and AHM and AFH were significantly different ($p < .001$). However, the techniques TS and AHM ($p > .05$), and RHM and AFH did not significantly differ. There was no significant interaction between the technique and the session ($F_{3,136} = 0.169$, $p > .05$).

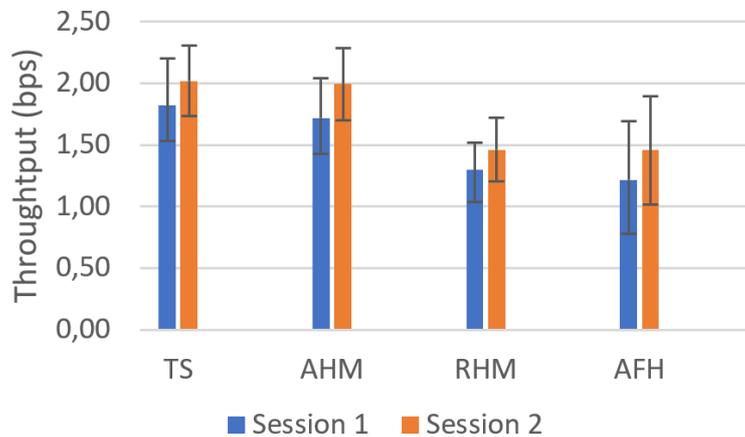

Figure 13: Throughput by session for different techniques

Figure 14 shows the error rate (%) by technique and by session. The rate is 2.62% (session1 $sd = 2.33$) and 1.77% (session 2 $sd = 1.83$) for the TS technique, 4.86% (session 1 $sd = 4.35$) and 4.94% (session 2 $sd = 3.20$) for the AHM technique, 8.33% (session 1 $sd = 7.07$) and 7.95% (session 2 $sd = 7.27$) for the RHM technique and 17.36% (session 1 $sd = 10.15$) and 12.81% (session 2 $sd = 6.42$) for the AFH technique.

A two-way 4 × 2 RM-ANOVA was carried out on error by session and techniques. The main effect of session on error rate was not significant ($F_{1,136} = 2.059$, $p > .05$). However, the main effect of technique on error rate was significant ($F_{3,136} = 31.251$, $p < .001$).

Tukey's HSD post hoc tests on techniques were carried out for error rate. Comparisons using the Tukey HSD test indicated that TS and RHM, TS and AFH, AHM and AFH, RHM and AFH are significantly different ($p < .001$). However, the techniques TS and AHM ($p > .05$), AHM and RHM did not significantly differ.

There was no significant interaction between the technique and the session ($F_{3,136} = 1.132$, $p > .05$).

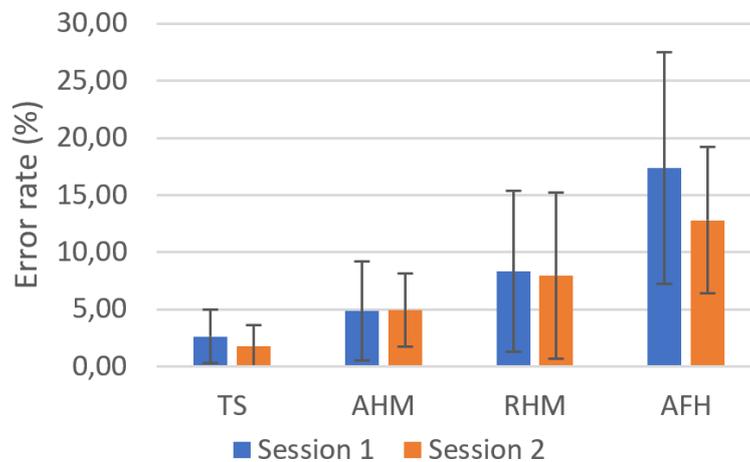

Figure 14: Error rates by session for different techniques

We then analyzed the results of the second session, which had a higher throughput compared to the first session for all techniques.

To test whether the four techniques follow Fitts' law, we separately examined the four different conditions of index of difficulty and computed the effective index of difficulty ($ID_e$) and the movement time for each condition.

We will use the Fitts' model based on the linear regression of the two variables Time Movement (MT) and the effective difficulty index ($ID_e$). The result is a regression equation of the form $MT = a + b \times ID_e$. Figure 15 and Table 3 show the regression line and the equation for each technique with R-squared values.

The R-squared values obtained for the models are above 0.8, so the models explain more than 80% of the variability of the data. Therefore, we conclude that the four techniques conform to Fitts' law.

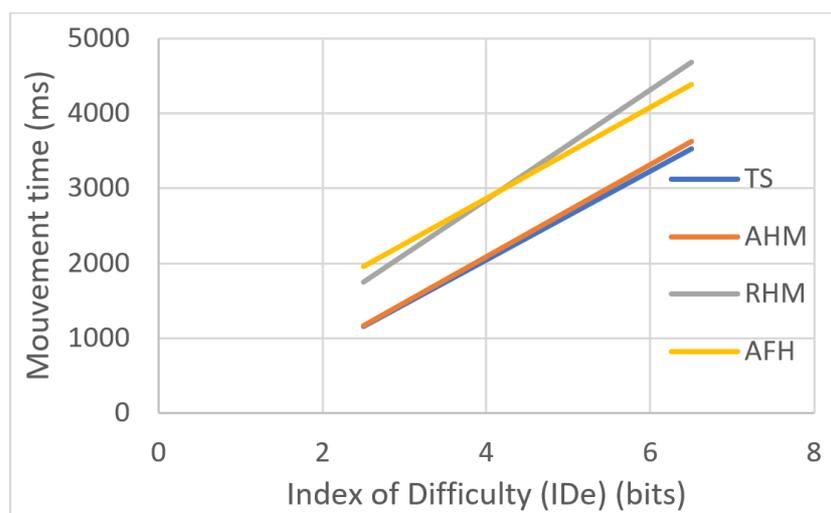

Figure 15: Fitts' law model for the four techniques (session 2)

Table 3: Regression equations are of the form $MT = a + b \times ID_e$

|  | Parameters | | |
| --- | --- | --- | --- |
|  | a | b | $r^2$ |
| TS | −326.47 | 592.71 | 0.923 |
| AHM | −361.57 | 613.58 | 0.993 |
| RHM | −81.762 | 732.09 | 0.927 |
| AFH | 437.08 | 607.5 | 0.834 |

Further analyses were conducted to examine the influence of target width on the error rates. Figure 16 shows the error rate for each target width W. The rate for $W = (100, 65, 50)$ is (0.31%, 2.16%, 4.32%) for the TS technique, (1.70%, 7.72%, 8.64%) for the AHM technique, (5.71%, 8.33%, 12.04%) for the RHM technique and (8.49%, 12.35%, 21.91%) for the AFH technique. A two-way 4 × 3 RM-ANOVA was carried out on errors by target width and techniques. The main effect of target width was significant ($F_{2,204} = 19.745, p < .001$). Also, the main effect of technique was significant $F = 3, 204 = 25.495, p < .001$).

Tukey's HSD post hoc tests on target width were carried out. Comparisons using the Tukey HSD test indicated that: difference between error rate with different target width is significant ($p < .05$).

There was no significant interaction between the target width and the technique ($F = 6, 204 = 1.907, p > .05$).

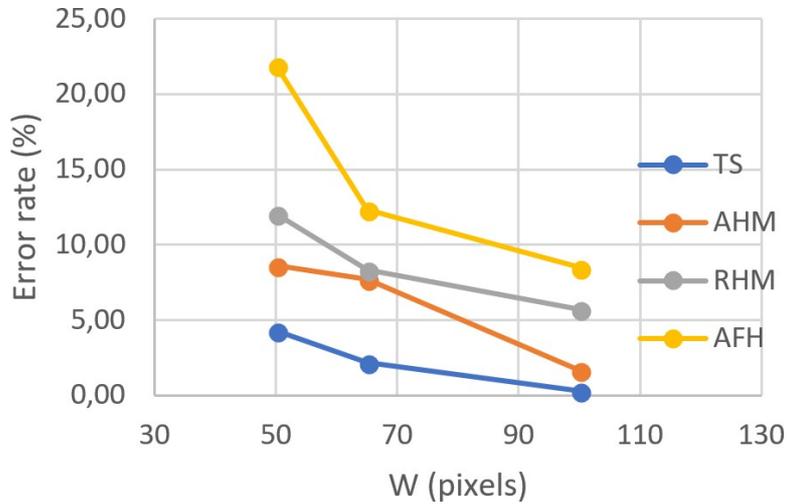

Figure 16: Error rates by target width (session 2)

## 7.2. Latency

The AFH technique is based on 3D tracking of the movement of the hand in a 3D environment. 3D input devices have tracking noise and latency [54]. This factor can degrade performance. The delay increases more because of the network communication used. We examined the system delay to ensure it is minimized and does not influence the results too much.

Lag, or latency, is the delay in the position updates of the apparatus [55]. It has already been studied in previous work on performance in 2D and 3D tasks [54, 56, 57, 58].

Mackenzie et al [56] formalize this lag and introduce it to equation (1) of linear regression as:

$$MT = a + (b + LAG) \times ID_e \quad (2)$$

The lag for the technique comes from two sources: the latency of the Fingo[2] camera (30$ms$ according to the technical documentation) and the latency related to our WIFI connection after multiple tests(10$ms$).

The predicted model for the AFH technique is:

$$MT = 437.08 + (607.5 + LAG) \times ID_e \quad (3)$$

With a $LAG = 0$, this reduces to:

$$MT = 437.08 + 567.5 \times ID_e \quad (4)$$

This can improve the throughput at 1.36$bps$ (compared to 1.34$bps$ with lag). Figure 17 shows the two equations of the AFH technique (the result found with LAG and the prediction without LAG). Lag is negligible in our study because it did not affect the results.

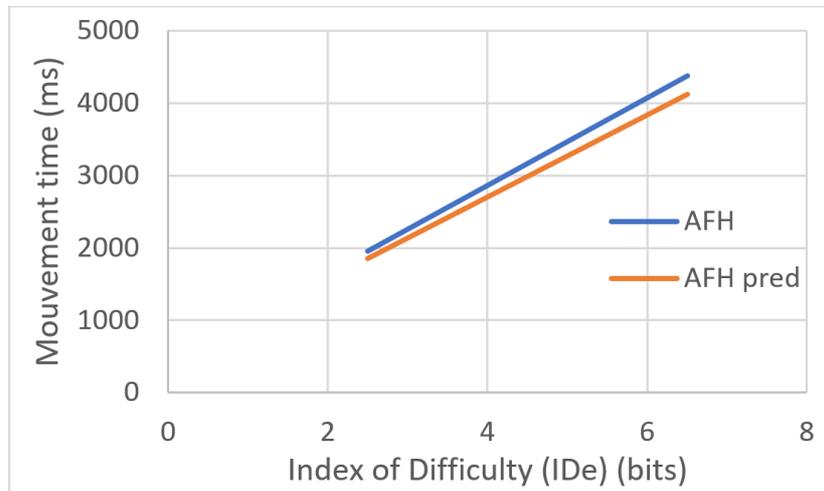

Figure 17: Fitts' law for AFH technique (found, predicted without lag) (session 2)

---

[2]https://www.usens.com/fingo

## 7.3. MacKenzie accuracy measures

Table 4 and Figure 18 show the means, standard deviations and Fisher's F value for the analysis of variance. For all measurements, there are significant differences among the four techniques. Also, for all measures, the lowest scores are the best, except for MO, where the score closest to 0 is the best. The units in Table 4 are "average number per selection task" for TRE, TAC, MDC and ODC; and "pixels" for MV, ME and MO.

Table 4: Means and standard deviations of accuracy measures for each technique.

| Variables | TS | | AHM | | RHM | | AFH | | F |
|---|---|---|---|---|---|---|---|---|---|
| | mean | sd | mean | sd | mean | sd | mean | sd | |
| Target re-entry (TRE) | 0.06 | 0.03 | 0.19 | 0.06 | 0.07 | 0.04 | 0.43 | 0.26 | 28.98 *** |
| Task Axis Crossing (TAC) | 0.88 | 0.15 | 1.37 | 0.11 | 0.72 | 0.17 | 2.33 | 0.63 | 82.30 *** |
| Movement Direction Change (MDC) | 1.68 | 0.17 | 3.29 | 0.45 | 1.41 | 0.26 | 5.35 | 1.44 | 99.90 *** |
| Orthogonal Direction Change (ODC) | 0.80 | 0.21 | 1.52 | 0.41 | 0.56 | 0.22 | 3.07 | 1.31 | 56.89 *** |
| Movement Variability (MV) | 28.88 | 4.54 | 24.78 | 4.43 | 30.89 | 6.89 | 25.99 | 2.74 | 5.81 ** |
| Movement Error (ME) | 32.89 | 3.90 | 26.65 | 4.48 | 48.65 | 9.68 | 24.97 | 1.79 | 63.41 *** |
| Movement Offset (MO) | -0.27 | 4.22 | -0.31 | 5.05 | -8.70 | 10.45 | 0.54 | 2.41 | 8.65 *** |

*** $p < .001$ ; ** $p < .01$

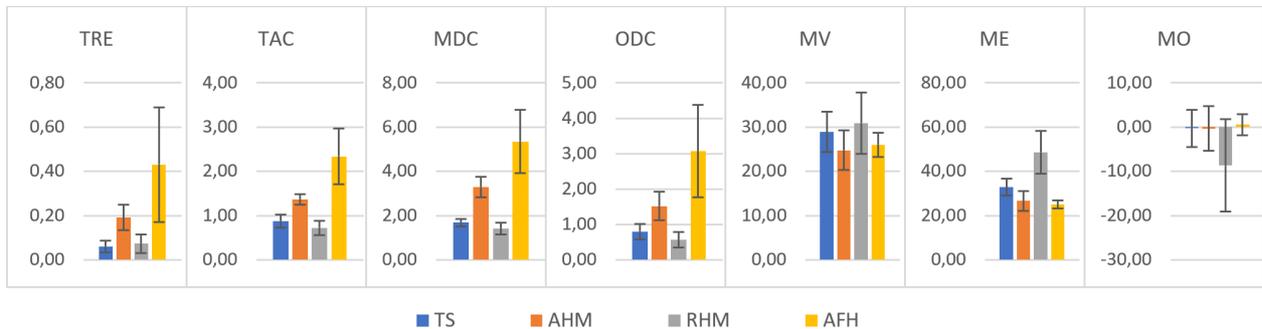

Figure 18: Accuracy measures for the four techniques

## 7.4. Participants feedback

### 7.4.1. SUS usability and participants preference

The results of the SUS questionnaires in Figure 19 indicate a better usability for the TS and AHM techniques with scores of 89.44 ($sd = 12.88$) and 85.69 ($sd = 9.92$), respectively, followed by the RHM technique with a score of 61.81 ($sd = 22.42$), and finally the AFH technique with a score of 47.50 ($sd = 23.46$).

A one-way 4 RM-ANOVA showed that the effect of the technique was significant ($F_{3,68} = 21.592, p < .001$). Tukey's HSD post hoc tests showed that the techniques are significantly different except between TS and AHM, and RHM and AFH ($p > .05$).

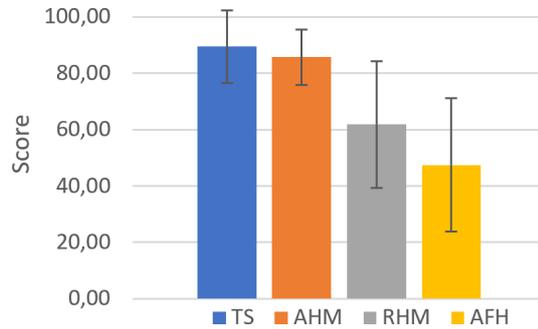

Figure 19: SUS usability score for the four techniques

We also asked participants to rank the technique in their preference order. Five participants preferred the TS technique, ten participants preferred the AHM technique, three participants preferred the RHM technique, and no one preferred the AFH technique. We assigned a score for each user preference position on a scale of 4 (4 for the most preferred technique of the user and 0 for the least preferred). Figure 20 shows the score received for each technique on a scale 100.

A one-way 4 RM-ANOVA showed that there is a significant difference between techniques ($F_{3,68}$ = 20.943, $p < .001$).

Tukey's HSD post hoc tests showed that the techniques are significantly different except between TS and AHM ($p > .05$).

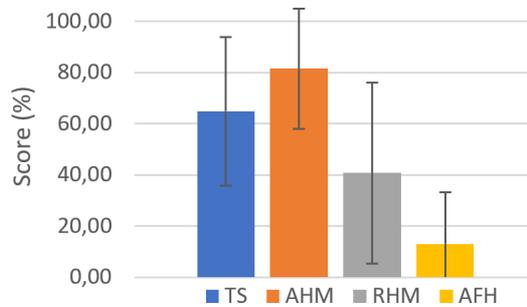

Figure 20: SUS usability score for the four techniques

### 7.4.2. Effort and comfort

The SUS questionnaire does not give a concrete indication of the weaknesses of the techniques and does not provide a precise basis of comparison between the techniques. However, to obtain more precise information on the techniques and to know how the participants felt about the use of these techniques, we collected the answers of the questionnaire derived from the ISO 9241-411 standard. The average obtained and the standard deviations for each question are shown in Figure 21. For all questions, there are significant differences among the four techniques. Since each answer was scored on a five-point scale, the value 3 corresponds to the midpoint and the values below 3 are negative. Only fatigue represents a significant difference between the two TS and AHM techniques according to the paired t-test.

The SUS questionnaire does not give a concrete indication of the weaknesses of the techniques and does not provide a precise basis of comparison between the techniques. However, to obtain more precise information on the techniques and to know how the participants felt about the use of these techniques, we collected the answers of the questionnaire derived from the ISO 9241-411 standard. The average obtained and the standard deviations for each question are shown in Figure 21. The value 3 corresponds to the midpoint and the values below 3 are negative.

A two-way 4 ×7 RM-ANOVA was carried out on different questions and techniques. The main effect questions were significant ($F_{6,476}$ = 5.359, $p < .001$). Also, the main effect of technique was significant ($F_{3,476}$ = 89.305, $p < .001$).

Tukey's HSD post hoc tests on techniques were carried out. Comparisons using the Tukey HSD test indicated that: TS and RHM, TS and AFH, AHM and AFH, AHM and RHM, RHM and AFH) are significantly different ($p < .001$). However, the techniques TS and AHM ($p > .05$) did not significantly differ.

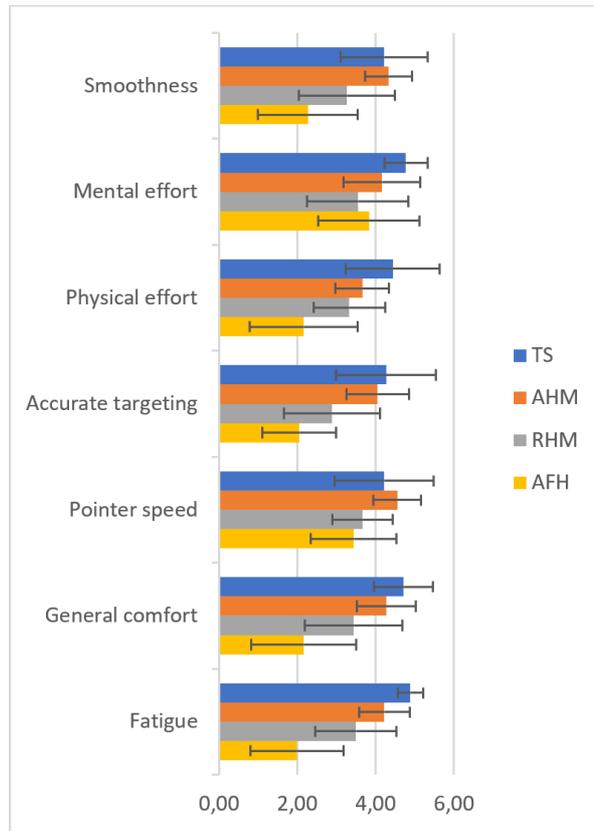

| Questions | TS | AHM | RHM | AFH |
|---|---|---|---|---|
| Smoothness | 4,22 | 4,33 | 3,28 | 2,28 |
| Mental effort | 4,78 | 4,17 | 3,46 | 3,83 |
| Physical effort | 4,44 | 3,67 | 3,33 | 2,17 |
| Accurate targeting | 4,28 | 4,06 | 2,89 | 2,06 |
| Pointer speed | 4,22 | 4,56 | 3,67 | 3,44 |
| General comfort | 4,72 | 4,28 | 3,44 | 2,17 |
| Fatigue | 4,89 | 4,22 | 3,50 | 2,00 |

Figure 21: Average score for the various comfort and effort questions for the four techniques. In all cases, a higher score is better (e.g., for physical and mental effort, a higher score indicates a lower effort

*7.4.3. Comments and observations*

The most common remark is about fatigue when using the AFH technique: *"arm discomfort after a while"*, *"for me, the movement with the free hand was the most complicated especially during a selection of small targets because the cursor was moving all the time"*, *"long-term free hand use, arm fatigue is felt"*, *"free hand very hard to keep calm"* and *"Discomfort during free-hand movement to keep your arm up for a long time"*. This could greatly affect the results in terms of time, errors, and usability.

For the RHM technique, the participants declare *"obliged to recalibrate often"*, *"the technique of the relative head movement is easily decalibrated"*.

However, users found that *"absolute head movement is fun to use"*, *"absolute head movement effective"*, *"The tactile surface and the absolute head movement were easy to use"*,

*"tactile surface a bit small but quite accurate and fast. Also, the absolute movement of the head is quite practical but still requires an adjustment on small targets"*.

In regard to the absolute head movement, a participant noted, *"It depends on the context of use I will choose the tactile surface or the absolute head movement"*.

Comments also on smart glasses: *"glasses poorly adapted to my head"*, *"by using the glasses, the eyes get tired. I do not usually wear glasses, the pressure on the edges of the nose is disturbing."*

Participants opinions were consistent with our observations during the experiment. Concerning the TS and AHM techniques, the participants performed the experiment very well. On the other hand, for the RHM technique, some participants were confused with the calibration and suggested that it was not intuitive. For the AFH technique, participants had some difficulties including fatigue, posture and hand position. Participants were required to keep their hands suspended in the air and, after some time, felt discomfort in the arm and the need to take a break. Participants needed to keep the index finger visible in front of the eyes (glasses and camera), for a better detection of its position, however, we noticed that some participants had the index finger very inclined and bent, which made it invisible to the camera, resulting in a loss of precision. Finally, some participants did not manage to position their index finger well in order to have the correct position of the cursor.

## 8. Discussion

The results show that there is a learning effect between the first and second sessions. This result was expected because the participants had never performed a pointing task on smart glasses prior to participating in this study. The movement time linearly increased with the target's index of difficulty for the four techniques.

In terms of throughput, the two techniques TS and AHM were generally the most efficient and demonstrated similar performances. The other two techniques, RHM and AFH, demonstrated the same performance without showing a significant difference. For all techniques, the error rates correlated with target width, as shown in Figure 16. The observed error rate was highest for the smallest target and gradually reduced with increasing size of the target.

The AFH technique correlated with faster transmission time but not for error rate. Figure 16 shows the error rate increases for the AFH technique when $W = 50$. The participants were fast but much less accurate than other techniques, especially for small targets, which penalizes this technique for accuracy. We also emphasize a high variance between the participants for this technique compared to other techniques.

For MacKenzie's precision measures, the most notable and obvious difference is between the AFH technique and other techniques for the first four measures (TRE, TAC, MDC, ODC), and AFH was the worst performer. We notice that AFH resulted in many inaccurate events in relation to the perfect pointing task, which correlates well with its low throughput. Hesitation and inaccuracy in selecting the target shown by the TRE measurement explain the significant error rate and the high "$a$" parameter of the regression equation, which represents the reaction time of the target selection. Participants were unable to place the cursor inside a target and hold it until the target was selected correctly. For the other three measures, the RHM technique stood out from the others. This resulted in a significant deviation of the cursor trajectory on the axis of the task which represents the perfect trajectory to have a low movement time (a high throughput). Participants were unable to hold the cursor on perfect trajectory because of difficulty with controlling movement speed. Thus, the movement time is high for this technique and its throughput remains rather low compared to the TS and AHM techniques.

TS and AHM techniques are the most efficient, but the TS technique is better than AHM on four measurements: TRE, TAC, MDC, ODC. The TRE measure explains the relatively higher error rate for AHM technique. This is not surprising because the participants had a high level of confidence controlling the cursor speed using the AHM technique.

The TAC, MDC, ODC measures show that there are directional changes for the AHM technique, unlike the TS technique for which the trajectory is rather straight. On the other hand, the AHM technique is more efficient than the TS technique on the three measurements: MV, ME, MO. This suggests that the participants manage to stay on the task axis more easily than a relative movement, but with difficulty controlling the speed of movement).

The results of the SUS questionnaire indicate a better usability for the TS and AHM techniques, followed by the RHM technique, and finally the AFH technique. These results align with the results of the other measures. Although the AHM technique was new to the participants, they found it very useful and preferred it to the TS technique (one that the participants had used previously).

For the ISO 9241-411 questionnaire the AFH technique is poorly evaluated against other techniques for all questions except the question of mental effort. The lowest scores were for fatigue, where participants reported feeling arm discomfort at the end of the experiment. They indicated that they had difficulty accurately selecting a target, which explains the error rate. The technique was not comfortable for them. The RHM technique was not as successful as the TS and AHM techniques with average scores, and demonstrated the worst rating for mental effort. Participants indicated that it was sometimes difficult to control the cursor with the position of the head for this technique.

## 9. Conclusion and perspectives

We explored four interaction techniques on smart glasses. All of them were functional. We conducted a user test experiment with 18 participants. The experiment aimed to demonstrate the strengths and weaknesses of these interaction techniques and to compare their use in conjunction with smart glasses.

The Tactile Surface (TS) and the Absolute Head Movement (AHM) techniques are the best in terms of performance and usability, but we found that the AHM technique is the best among the techniques that do not require a handheld device. The Absolute Free Hand (AFH) technique is very promising in realistic scenarios where interaction would happen only occasionally or not so intensively. Many researchers and companies are devoting time and effort to this area of user experience research. According to the report published by MarketsandMarkets, novel healthcare application development is expected to emerge as a significant market for free hand technologies in 2022 [59]. As we evaluated these techniques only for pointing at a target, and not for other tasks, future studies are needed to explore the limitations of accuracy or high effort that we were exposed to with the tasks presented here.

In our experiment, the TS technique was the baseline for the head movement and absolute free hand techniques. TS was the most common prior technique for user, but unfortunately not the most suitable in a healthcare context. Overall, we expect to see a wide range of input methods for smart glasses.

Future work is needed to optimize the Absolute Free Hand technique. We also plan to investigate a new validation technique without using a button to click on targets. The goal is pointing without carrying an additional device. The validation technique must also be studied, because using a button on a remote device is not suitable in a context where users need freedom with their hands. The next step is to study the pointing technique in the healthcare context.

## 10. Acknowledgments

This work pursues the eGLASSES project, which was partially funded by NCBiR, FWF, SNSF, ANR and FNR under the ERA-NET CHIST-ERA II framework. The authors thank all the volunteers for the participation in the experience

## References

[1] I. Pecci, B. Martin, I. Kacem, I. Maamria, S. Faye, N. Louveton, G. Gheorghe, T. Engel, Not a tile out of place: Toward creating context-dependent user interfaces on smartglasses, in: Human System Interactions (HSI), 2016 9th International Conference on, IEEE, 2016, pp. 497–503, DOI: http://dx.doi.org/10.1109/HSI.2016.7529680.

[2] Google glass project, https://en.wikipedia.org/wiki/Google-Glass/, retrieved May 20, 2018.

[3] S. Yi, Z. Qin, E. Novak, Y. Yin, Q. Li, Glassgesture: Exploring head gesture interface of smart glasses, in: Computer Communications, IEEE INFOCOM 2016-The 35th Annual IEEE International Conference on, IEEE, 2016, pp. 1–9, DOI: http://dx.doi.org/10.1109/INFOCOM.2016.7524542.

[4] B. Kollee, S. Kratz, A. Dunnigan, Exploring gestural interaction in smart spaces using head mounted devices with ego-centric sensing, in: Proceedings of the 2nd ACM symposium on Spatial user interaction, ACM, 2014, pp. 40–49, DOI: http://dx.doi.org/10.1145/2659766.2659781.

[5] L.-H. Lee, P. Hui, Interaction methods for smart glasses: A survey, IEEE Access.

[6] S. Mitrasinovic, E. Camacho, N. Trivedi, J. Logan, C. Campbell, R. Zilinyi, B. Lieber, E. Bruce, B. Taylor, D. Martineau, et al., Clinical and surgical applications of smart glasses, Technology and Health Care 23 (4) (2015) 381–401, DOI: http://dx.doi.org/10.3233/THC-150910.

[7] M. Prgomet, A. Georgiou, J. I. Westbrook, The impact of mobile handheld technology on hospital physicians' work practices and patient care: a systematic review, Journal of the American Medical Informatics Association 16 (6) (2009) 792–801, DOI: http://dx.doi.org/10.1197/jamia.M3215.

[8] M. Göken, A. N. Başoğlu, M. Dabic, Exploring adoption of smart glasses: Applications in medical industry, in: Management of Engineering and Technology (PICMET), 2016 Portland International Conference on, IEEE, 2016, pp. 3175–3184, DOI: http://dx.doi.org/10.1109/PICMET.2016.7806835.

[9] Augmedix company, https://www.augmedix.com/about/, retrieved May 20, 2018.

[10] C. Plaisant, R. Mushlin, A. Snyder, J. Li, D. Heller, B. Shneiderman, Lifelines: using visualization to enhance navigation and analysis of patient records, in: The Craft of Information Visualization, Elsevier, 2003, pp. 308–312, DOI: http://dx.doi.org/10.1016/B978-155860915-0/50038-X.

[11] A. Faiola, C. Newlon, Advancing critical care in the icu: a human-centered biomedical data visualization systems, in: International Conference on Ergonomics and Health Aspects of Work with Computers, Springer, 2011, pp. 119–128, DOI: http://dx.doi.org/10.1007/978-3-642-21716-6_13.

[12] I. Belkacem, I. Pecci, B. Martin, L'exploration d'un espace de conception pour la visualisation des dossiers patients, in: 29ème conférence francophone sur l'Interaction Homme-Machine, 2017, pp. 10–p, DOI: http://dx.doi.org/10.1145/3132129.3132165.

[13] Y.-C. Tung, C.-Y. Hsu, H.-Y. Wang, S. Chyou, J.-W. Lin, P.-J. Wu, A. Valstar, M. Y. Chen, User-defined game input for smart glasses in public space, in: Proceedings of the 33rd Annual ACM Conference on Human Factors in Computing Systems, ACM, 2015, pp. 3327–3336, DOI: http://dx.doi.org/10.1145/2702123.2702214.

[14] M. F. Roig-Maimó, I. S. MacKenzie, C. Manresa-Yee, J. Varona, Head-tracking interfaces on mobile devices: Evaluation using fitts law and a new multi-directional corner task for small displays, International Journal of Human-Computer Studies 112 (2018) 1–15, DOI: http://dx.doi.org/10.1016/j.ijhcs.2017.12.003.

[15] M. Ward, R. Azuma, R. Bennett, S. Gottschalk, H. Fuchs, A demonstrated optical tracker with scalable work area for head-mounted display systems, in: Proceedings of the 1992 symposium on Interactive 3D graphics, ACM, 1992, pp. 43–52, DOI: http://dx.doi.org/10.1145/147156.147162.

[16] M. Bichsel, A. Pentland, Automatic interpretation of human head movements, in: 13th International Joint Conference on Artificial Intelligence (IJCAI), Workshop on Looking At People, Chambery France, 1993.

[17] M. B. López, J. Hannuksela, O. Silvén, L. Fan, Head-tracking virtual 3-d display for mobile devices, in: Computer Vision and Pattern Recognition Workshops (CVPRW), 2012 IEEE Computer Society Conference on, IEEE, 2012, pp. 27–34, DOI: http://dx.doi.org/10.1109/CVPRW.2012.6238891.

[18] C. J. Lin, S.-H. Ho, Y.-J. Chen, An investigation of pointing postures in a 3d stereoscopic environment, Applied ergonomics 48 (2015) 154–163, DOI: http://dx.doi.org/10.1016/j.apergo.2014.12.001.

[19] A. M. Bernardos, D. Gómez, J. R. Casar, A comparison of head pose and deictic pointing interaction methods for smart environments, International Journal of Human-Computer Interaction 32 (4) (2016) 325–351, DOI: http://dx.doi.org/10.1080/10447318.2016.1142054.

[20] C. A. Avizzano, P. Sorace, D. Checcacci, M. Bergamasco, A navigation interface based on head tracking by accelerometers, in: Robot and Human Interactive Communication, 2004. ROMAN 2004. 13th IEEE International Workshop on, IEEE, 2004, pp. 625–630, DOI: http://dx.doi.org/10.1109/ROMAN.2004.1374834.

[21] Y. Chi, S. Ong, M. Yuan, A. Nee, Wearable interface for the physical disabled, in: Proceedings of the 1st international convention on Rehabilitation engineering & assistive technology: in conjunction with 1st Tan Tock Seng Hospital Neurorehabilitation Meeting, ACM, 2007, pp. 28–32, DOI: http://dx.doi.org/10.1145/1328491.1328500.

[22] R. Malkewitz, Head pointing and speech control as a hands-free interface to desktop computing, in: Proceedings of the third international ACM conference on Assistive technologies, ACM, 1998, pp. 182–188, DOI: http://dx.doi.org/10.1145/274497.274531.

[23] G.-M. Eom, K.-S. Kim, C.-S. Kim, J. Lee, S.-C. Chung, B. Lee, H. Higa, N. Furuse, R. Futami, T. Watanabe, Gyro-mouse for the disabled, International Journal of Control, Automation, and Systems 5 (2) (2007) 147–154.

[24] R. Atienza, R. Blonna, M. I. Saludares, J. Casimiro, V. Fuentes, Interaction techniques using head gaze for virtual reality, in: Region 10 Symposium (TENSYMP), 2016 IEEE, IEEE, 2016, pp. 110–114, DOI: http://dx.doi.org/10.1109/TENCONSpring.2016.7519387.

[25] F. Wahl, M. Freund, O. Amft, Using smart eyeglasses as a wearable game controller, in: Adjunct Proceedings of the 2015 ACM International Joint Conference on Pervasive and Ubiquitous Computing and Proceedings of the 2015 ACM International Symposium on Wearable Computers, ACM, 2015, pp. 377–380, DOI: http://dx.doi.org/10.1145/2800835.2800914.

[26] C. Yu, Y. Gu, Z. Yang, X. Yi, H. Luo, Y. Shi, Tap, dwell or gesture?: Exploring head-based text entry techniques for hmds, in: Proceedings of the 2017 CHI Conference on Human Factors in Computing

Systems, ACM, 2017, pp. 4479–4488, DOI: http://dx.doi.org/10.1145/3025453.3025964.
[27] S. Jalaliniya, D. Mardanbeigi, T. Pederson, D. W. Hansen, Head and eye movement as pointing modalities for eyewear computers, in: Wearable and Implantable Body Sensor Networks Workshops (BSN Workshops), 2014 11th International Conference on, IEEE, 2014, pp. 50–53, DOI: http://dx.doi.org/10.1109/BSN.Workshops.2014.14.
[28] S. Jalaliniya, D. Mardanbegi, T. Pederson, Magic pointing for eyewear computers, in: Proceedings of the 2015 ACM International Symposium on Wearable Computers, ACM, 2015, pp. 155–158, DOI: http://dx.doi.org/10.1145/2802083.2802094.
[29] M. Kytö, B. Ens, T. Piumsomboon, G. A. Lee, M. Billinghurst, Pinpointing: Precise head-and eye-based target selection for augmented reality, in: Proceedings of the 2018 CHI Conference on Human Factors in Computing Systems, ACM, 2018, p. 81, DOI: http://dx.doi.org/10.1145/3173574.3173655.
[30] Y. Y. Qian, R. J. Teather, The eyes don't have it: an empirical comparison of head-based and eye-based selection in virtual reality, in: Proceedings of the 5th Symposium on Spatial User Interaction, ACM, 2017, pp. 91–98, DOI: http://dx.doi.org/10.1145/3131277.3132182.
[31] Gaze input, Microsoft HoloLens, https://docs.microsoft.com/en-us/windows/mixed-reality/gaze, retrieved May 20, 2018.
[32] M. Kolsch, M. Turk, T. Hollerer, Vision-based interfaces for mobility, in: Mobile and Ubiquitous Systems: Networking and Services, 2004. MOBIQUITOUS 2004. The First Annual International Conference on, IEEE, 2004, pp. 86–94, DOI: http://dx.doi.org/10.1109/MOBIQ.2004.1331713.
[33] T. Lee, T. Hollerer, Handy ar: Markerless inspection of augmented reality objects using fingertip tracking, in: Wearable Computers, 2007 11th IEEE International Symposium on, IEEE, 2007, pp. 83–90, DOI: http://dx.doi.org/10.1109/ISWC.2007.4373785.
[34] G. Heo, D.-W. Lee, H.-C. Shin, H.-T. Jeong, T.-W. Yoo, Hand segmentation and fingertip detection for interfacing of stereo vision-based smart glasses, in: Consumer Electronics (ICCE), 2015 IEEE International Conference on, IEEE, 2015, pp. 585–586, DOI: http://dx.doi.org/10.1109/ICCE.2015.7066537.
[35] L. Chan, Y.-L. Chen, C.-H. Hsieh, R.-H. Liang, B.-Y. Chen, Cyclopsring: Enabling wholehand and context-aware interactions through a fisheye ring, in: Proceedings of the 28th Annual ACM Symposium on User Interface Software & Technology, ACM, 2015, pp. 549–556, DOI: http://dx.doi.org/10.1145/2807442.2807450.
[36] D. Kim, O. Hilliges, S. Izadi, A. D. Butler, J. Chen, I. Oikonomidis, P. Olivier, Digits: freehand 3d interactions anywhere using a wrist-worn gloveless sensor, in: Proceedings of the 25th annual ACM symposium on User interface software and technology, ACM, 2012, pp. 167–176, DOI: http://dx.doi.org/10.1145/2380116.2380139.
[37] G. Bailly, J. Müller, M. Rohs, D. Wigdor, S. Kratz, Shoesense: a new perspective on gestural interaction and wearable applications, in: Proceedings of the SIGCHI Conference on Human Factors in Computing Systems, ACM, 2012, pp. 1239–1248, DOI: http://dx.doi.org/10.1145/2207676.2208576.
[38] Z. Huang, W. Li, P. Hui, Ubii: Towards seamless interaction between digital and physical worlds, in: Proceedings of the 23rd ACM international conference on Multimedia, ACM, 2015, pp. 341–350, DOI: http://dx.doi.org/10.1145/2733373.2806266.
[39] T. Ha, S. Feiner, W. Woo, Wearhand: Head-worn, rgb-d camera-based, bare-hand user interface with visually enhanced depth perception, in: Mixed and Augmented Reality (ISMAR), 2014 IEEE International Symposium on, IEEE, 2014, pp. 219–228, DOI: http://dx.doi.org/10.1109/ISMAR.2014.6948431.
[40] M. S. M. Y. Sait, S. P. Sargunam, D. T. Han, E. D. Ragan, Physical hand interaction for controlling multiple virtual objects in virtual reality, in: Proceedings of the 3rd International Workshop on Interactive and Spatial Computing, ACM, 2018, pp. 64–74, DOI: http://dx.doi.org/10.1145/3191801.3191814.
[41] S. Gustafson, D. Bierwirth, P. Baudisch, Imaginary interfaces: spatial interaction with empty hands and without visual feedback, in: Proceedings of the 23nd annual ACM symposium on User interface software and technology, ACM, 2010, pp. 3–12, DOI: http://dx.doi.org/10.1145/1866029.1866033.

[42] A. Colaço, A. Kirmani, H. S. Yang, N.-W. Gong, C. Schmandt, V. K. Goyal, Mime: compact, low power 3d gesture sensing for interaction with head mounted displays, in: Proceedings of the 26th annual ACM symposium on User interface software and technology, ACM, 2013, pp. 227–236, DOI: http://dx.doi.org/10.1145/2501988.2502042.

[43] Gesture input, Microsoft HoloLens, https://docs.microsoft.com/en-us/windows/mixed-reality/gestures, retrieved May 20, 2018.

[44] Meta 2, https://meta-eu.myshopify.com/, retrieved May 20, 2018.

[45] E. N. Arcoverde Neto, R. M. Duarte, R. M. Barreto, J. P. Magalhães, C. Bastos, T. I. Ren, G. D. Cavalcanti, Enhanced real-time head pose estimation system for mobile device, Integrated Computer-Aided Engineering 21 (3) (2014) 281–293, DOI: http://dx.doi.org/10.3233/ICA-140462.

[46] K. Andersson, Manipulating control-display ratios in room-scale virtual reality (2017).

[47] P. M. Fitts, The information capacity of the human motor system in controlling the amplitude of movement., Journal of experimental psychology 47 (6) (1954) 381, DOI: http://dx.doi.org/10.1037/h0055392.

[48] I. S. MacKenzie, A note on the information-theoretic basis for fitts law, Journal of motor behavior 21 (3) (1989) 323–330, DOI: http://dx.doi.org/10.1080/00222895.1989.10735486.

[49] A. T. Welford, Fundamentals of skill.

[50] R. W. Soukoreff, I. S. MacKenzie, Towards a standard for pointing device evaluation, perspectives on 27 years of fitts law research in hci, International journal of human-computer studies 61 (6) (2004) 751–789, DOI: http://dx.doi.org/10.1016/j.ijhcs.2004.09.001.

[51] . ISO, Ergonomics of human-system interaction part 411: Evaluation methods for the design of physical input devices.

[52] S. Zhai, Characterizing computer input with fitts law parametersthe information and non-information aspects of pointing, International Journal of Human-Computer Studies 61 (6) (2004) 791–809, DOI: http://dx.doi.org/10.1016/j.ijhcs.2004.09.006.

[53] I. S. MacKenzie, T. Kauppinen, M. Silfverberg, Accuracy measures for evaluating computer pointing devices, in: Proceedings of the SIGCHI conference on Human factors in computing systems, ACM, 2001, pp. 9–16, DOI: http://dx.doi.org/10.1145/365024.365028.

[54] R. J. Teather, A. Pavlovych, W. Stuerzlinger, I. S. MacKenzie, Effects of tracking technology, latency, and spatial jitter on object movement, in: 3D User Interfaces, 2009. 3DUI 2009. IEEE Symposium on, IEEE, 2009, pp. 43–50, DOI: http://dx.doi.org/10.1109/3DUI.2009.4811204.

[55] E. Foxlin, et al., Motion tracking requirements and technologies, Handbook of virtual environment technology 8 (2002) 163–210.

[56] I. S. MacKenzie, C. Ware, Lag as a determinant of human performance in interactive systems, in: Proceedings of the INTERACT'93 and CHI'93 conference on Human factors in computing systems, ACM, 1993, pp. 488–493, DOI: http://dx.doi.org/10.1145/169059.169431.

[57] R. H. So, G. K. Chung, Sensory motor responses in virtual environments: Studying the effects of image latencies for target-directed hand movement, in: Engineering in Medicine and Biology Society, 2005. IEEE-EMBS 2005. 27th Annual International Conference of the, IEEE, 2005, pp. 5006–5008, DOI: http://dx.doi.org/10.1109/IEMBS.2005.1615599.

[58] C. Ware, R. Balakrishnan, Reaching for objects in vr displays: lag and frame rate, ACM Transactions on Computer-Human Interaction (TOCHI) 1 (4) (1994) 331–356, DOI: http://dx.doi.org/10.1145/198425.198426.

[59] Gesture recognition and touchless sensing market by technology (touch-based and touchless), product (sanitary equipment, touchless biometric), industry, and geography - global forecast to 2022, marketsandmarkets.com,, http://www.marketsandmarkets.com/Market-Reports/touchless-sensinggesturing-market-369.html/, retrieved December 20, 2018.